\documentclass[conference]{IEEEtran}
\IEEEoverridecommandlockouts
\usepackage{algorithm}
\usepackage{algorithmic}
\usepackage{flushend}
\usepackage{cite}
\usepackage{amsmath,amssymb,amsfonts}
\usepackage{algorithmic}
\usepackage{graphicx}
\usepackage{textcomp}
\usepackage{xcolor}
\def\BibTeX{{\rm B\kern-.05em{\sc i\kern-.025em b}\kern-.08em
    T\kern-.1667em\lower.7ex\hbox{E}\kern-.125emX}}
\begin{document}

\title{Joint Space-Time Adaptive Processing and Beamforming Design for Cell-Free ISAC Systems\\
\thanks{This work was supported in part by the U.S. National Science Foundation under Grant CCF-2225575 and Grant CCF-2322191, in part by the National Natural Science Foundation of China under Grant 62371090 and Grant 62471086,  in part by Fundamental Research Funds for the Central Universities (Grant No. DUT24ZD125 and DUT24RC(3)005), and in part by Liaoning Applied Basic Research Program (2023JH2/101700364).}
}

\author{
    \IEEEauthorblockN{Rang Liu$^{\star}$, Ming Li$^{\dagger}$, and Qian Liu$^{\dagger}$}
    \IEEEauthorblockA{
       $^{\star}$University of California, Irvine, CA 92697, USA \\
        $^{\dagger}$Dalian University of Technology, Dalian, Liaoning 116024, China \\
        Email: \texttt{rangl2@uci.edu, mli@dlut.edu.cn, qianliu@dlut.edu.cn}
    }
}


\maketitle

\begin{abstract}
In this paper, we explore cooperative sensing and communication within cell-free integrated sensing and communication (ISAC) systems. Specifically, multiple transmit access points (APs) collaboratively serve multiple communication users while simultaneously illuminating a potential target, with a separate sensing AP dedicated to collecting echo signals for target detection. To improve the performance of identifying a moving target in the presence of strong interference originating from transmit APs, we employ the space-time adaptive processing (STAP) technique and jointly optimize the transmit/receive beamforming. Our goal is to maximize the radar output signal-to-interference-plus-noise ratio (SINR), subject to constraints on the communication SINR and transmit power. An efficient algorithm is developed to solve the resulting non-convex optimization problem. Simulations demonstrate significant performance improvements in target detection and validate the advantages of the proposed joint STAP and beamforming design for cell-free ISAC systems.
\end{abstract}


\section{Introduction}
Integrated sensing and communication (ISAC) has been recognized as one of the key usage scenarios for the sixth-generation (6G) wireless networks.  
ISAC facilitates the sharing of spectral resources, hardware platforms, and signal processing modules between sensing and communication (S\&C) functions, enabling significantly higher efficiencies \cite{Zhang-ICST-2022, Liu JSAC 2022}.
Although advanced beamforming/waveform designs have been extensively investigated to enhance S\&C performance for monostatic ISAC systems \cite{Liu JSTSP 2021,Zhang-JSTSP-2021, Liu-WCM-2023, Xiao TCOM 2024, Li TWC 2024}, achieving reliable performance in complex environments remains a challenge due to factors such as limited service coverage, significant self-interference, and limitations of a single observation perspective. To address these challenges, there is growing interest in cooperative S\&C approaches for cell-free networks \cite{Wei Netw 2024, Wei TVT 2024, Meng arxiv1 2024}.

By leveraging geographic diversity and the multiple perspectives provided by widely distributed access points (APs), both S\&C performance in cell-free ISAC systems can be greatly improved through advanced beamforming designs \cite{Gao JSAC 2024, Elfiatoure Globecom 2023, Rivetti arxiv 2024, Meng arxiv 2024, Demirhan arxiv 2024, Buzzi ISWCS 2024, Elfiatoure arxiv 2024, Liu WCL 2024, Behdad TWC 2024, Liu WCNC 2024, Wang arxiv 20224}. The authors in \cite{Demirhan arxiv 2024} focused on a scenario with multiple transmit APs and a single sensing AP, aiming to maximize the sensing signal-to-noise ratio (SNR) while satisfying the communication signal-to-interference-plus-noise ratio (SINR) requirements. The scalability of such systems was addressed in \cite{Buzzi ISWCS 2024}, where the authors considered setups involving multiple transmit and receive APs. In addition, a multi-target scenario was explored in \cite{Elfiatoure arxiv 2024}. While these works investigated beamforming designs for various cell-free ISAC applications \cite{Gao JSAC 2024, Elfiatoure Globecom 2023, Rivetti arxiv 2024, Meng arxiv 2024, Demirhan arxiv 2024, Buzzi ISWCS 2024, Elfiatoure arxiv 2024}, they overlooked the critical issue of interference between transmit and receive APs, which is particularly problematic in scenarios with dense AP deployments. To address this gap, the authors in \cite{Liu WCL 2024, Behdad TWC 2024, Liu WCNC 2024, Wang arxiv 20224} incorporated this interference into their optimization frameworks and developed efficient beamforming algorithms to suppress it. However, unlike common multi-user interference in communication systems, the interference that propagates directly from transmit APs to the sensing APs manifests as signal-dependent clutter, posing challenges for suppression using conventional spatial beamforming techniques.
Moreover, the more complex scenario involving a moving target, in which different paths are associated with varying propagation delays and Doppler shifts, has not yet been explored.

Motivated by the above, in this paper we focus on detecting a moving target in the presence of strong interference originating from transmit APs. A general sensing model incorporating propagation delays and Doppler shifts is established for the considered cell-free ISAC systems, in which multiple transmit APs cooperatively communicate with multiple users and simultaneously probe a moving target, and one sensing AP collects echo signals to perform target detection. To achieve satisfactory clutter suppression and target detection performance, we employ the space-time adaptive processing (STAP) technique \cite{Liu-JSAC-2022}, \cite{Liu-JSTSP-2022}. The beamforming matrices at the transmit APs and the space-time filter at the sensing AP are jointly optimized to maximize the radar SINR as well as satisfy the communication SINR requirements and power budget. 
Simulation results demonstrate the superiority of the proposed joint STAP and beamforming design algorithm compared to approaches that either lack receive beamforming or utilize spatial beamforming alone.

\section{System Model}
\label{Sec:System Model}

We consider a cooperative cell-free ISAC system composed of $B$ transmit APs, which transmit dual-functional signals to cooperatively serve $K$ communication users and simultaneously illuminate one point-like moving target, and one sensing AP receives echo signals to identify the presence or absence of this target. 
Without loss of generality, we assume that each transmit AP is equipped with $N_\text{t}$ antennas and the receive AP is equipped with $N_\text{r}$ antennas, all of which are arranged in uniform linear arrays (ULAs) with half-wavelength antenna spacing. 
To improve target detection performance in the presence of strong clutters, particularly those originating directly from the transmitters to the sensing receiver, we propose to utilize the STAP technique and jointly optimize the transmit beamforming and the receive filter in both the spatial and temporal domains.

The transmitted dual-functional signal from the $b$-th transmit AP at the $l$-th time slot is expressed as
\begin{equation}
    \mathbf{x}_{b}[l] = \mathbf{W}_{\text{c},b}\mathbf{s}_{\text{c}}[l] + \mathbf{W}_{\text{r},b}\mathbf{s}_{\text{r}}[l] = \mathbf{W}_b\mathbf{s}[l],
\end{equation}
where $\mathbf{s}_\text{c}[l]\in\mathbb{C}^K$ denotes the communication symbols that are precoded by the beamforming matrix $\mathbf{W}_{\text{c},b}\in\mathbb{C}^{N_\text{t}\times K}$, and $\mathbf{s}_\text{r}[l]\in\mathbb{C}^{N_\text{t}}$ denotes the radar probing symbols that are precoded by the beamforming matrix $\mathbf{W}_{\text{r},b}\in\mathbb{C}^{N_\text{t}\times N_\text{t}}$. For simplicity, we define $\mathbf{s}[l]\triangleq [\mathbf{s}^T_\text{c}[l]~\mathbf{s}^T_\text{r}[l]]^T$ and $\mathbf{W}_b\triangleq [\mathbf{W}_{\text{c},b}~\mathbf{W}_{\text{r},b}]$. The transmitted symbols are assumed to be statistically independent, i.e., $\mathbb{E}\{\mathbf{s}[l]\mathbf{s}^H[l]\} = \mathbf{I}_{N_\text{t}+K}$.

For downlink multiuser communications, the received signal at the $k$-th user can be written as 
\begin{equation}
    y_k[l] = \sum_{b=1}^B\mathbf{h}^T_{b,k}\mathbf{x}_{b}[l] + n_{k}[l],
\end{equation}
where $\mathbf{h}_{b,k}\in\mathbb{C}^{N_\text{t}}$ represents the channel between the $b$-th transmit AP and the $k$-th user, and $n_k[l]\sim\mathcal{CN}(0,\sigma^2_\text{c})$ is the additive white Gaussian noise (AWGN). The SINR for the $k$-th user is calculated as 
\begin{equation}
    \text{SINR}_{\text{c},k} = \frac{\big|\sum_{b=1}^B\mathbf{h}^T_{b,k}\mathbf{w}_{b,k}\big|^2}{\sum_{j=1,j\neq k}^{K+N_\text{t}}\big|\sum_{b=1}^B\mathbf{h}^T_{b,k}\mathbf{w}_{b,j}\big|^2 + \sigma^2_\text{c}},
\end{equation}
where $\mathbf{w}_{b,j}$ represents the $j$-th column of $\mathbf{W}_b$.

The transmitted dual-functional signals reach the target and are subsequently reflected back to the sensing AP. The baseband target echo signal at the sensing AP is
\begin{equation}
    \mathbf{y}_\text{t}[l] = \sum_{b=1}^B\alpha_b\mathbf{a}_\text{r}(\theta_\text{t})\mathbf{a}^T_\text{t}(\theta_{b,\text{t}})\mathbf{x}_b[l-\tau_b]e^{\jmath2\pi (l-1)f_{\text{D},b}},
\end{equation}
where $\alpha_b\sim\mathcal{CN}(0,\sigma^2_b)$ denotes the complex channel gain composed of the radar cross section (RCS) of the target and the distance-dependent path loss when the illuminating signal comes from the $b$-th transmit AP, $\mathbf{a}_\text{r}(\theta_\text{t})\in\mathbb{C}^{N_\text{r}}$ and $\mathbf{a}_\text{t}(\theta_{b,\text{t}})\in\mathbb{C}^{N_\text{t}}$ are steering vectors, $\theta_\text{t}$ is the angle of the target with respect to the sensing AP, and $\theta_{b,\text{t}}$ is the angle of the target with respect to the $b$-th transmit AP. The parameters $\tau_b$ and $f_{\text{D},b}$ represent the propagation delay and Doppler shift associated with the link of the $b$-th transmit AP -- target -- sensing AP, which are respectively quantized as integers based on the sampling interval and frequency. 
Since the signals from the transmit APs can also directly propagate to the sensing AP, the received signals at the sensing AP are composed of the target echos $\mathbf{y}_\text{t}[l]$, the clutters from the transmit APs and the noise, which is expressed as 
\begin{equation}
    \mathbf{y}_\text{r}[l] = \mathbf{y}_\text{t}[l]+ \sum_{b=1}^B\mathbf{G}_b\mathbf{x}_b[l-\iota_b] + \mathbf{n}_\text{r}[l],
\end{equation}
where $\mathbf{G}_b\in\mathbb{C}^{N_\text{r}\times N_\text{t}}$ denotes the channel between the $b$-th transmit AP and the sensing AP, $\iota_b$ is the propagation delay from the $b$-th transmit AP to the sensing AP, and $\mathbf{n}_\text{r}[l]\sim\mathcal{CN}(\mathbf{0},\sigma^2_\text{r}\mathbf{I}_{N_\text{r}})$ is the AWGN. 
After transmitting $L$ symbols, the received signals collected during $Q \geq L+\max\{\tau_b,~\forall b\}$ time slots,  which is defined as $\mathbf{Y}_\text{r}\triangleq[\mathbf{y}_\text{r}[1], \mathbf{y}_\text{r}[2],\ldots,\mathbf{y}_\text{r}[Q]]$, can be written as 
\begin{equation}
    \mathbf{Y}_\text{r} = \sum_{b=1}^B\alpha_b\mathbf{a}_\text{r}(\theta_\text{t})\mathbf{a}^T_\text{t}(\theta_{b,\text{t}})\mathbf{X}_b\mathbf{J}_{\tau_b}\mathbf{D}_b+ \sum_{b=1}^B\mathbf{G}_b\mathbf{X}_b\mathbf{J}_{\iota_b} + \mathbf{N}_\text{r},
\end{equation}
where the signal matrix $\mathbf{X}_b\triangleq[\mathbf{x}_b[1], \mathbf{x}_b[2],\ldots,\mathbf{x}_b[L]]$ associated with the symbol matrix $\mathbf{S}\triangleq[\mathbf{s}[1], \mathbf{s}[2],\ldots,\mathbf{s}[L]]$, the shift matrix $\mathbf{J}_{\tau_b}\in\mathbb{R}^{L\times Q}$ where the $(m,n)$-th element equals 1 if $m-n +\tau_b = 0$ and otherwise 0, the shift matrix $\mathbf{J}_{\iota_b}\in\mathbb{R}^{L\times Q}$ is defined in the same way, the Doppler response matrix $\mathbf{D}_b\triangleq\text{diag}\{\mathbf{d}_b(f_{\text{D},b})\}$ with $\mathbf{d}_b(f_{\text{D},b})\triangleq [1, e^{\jmath 2\pi f_{\text{D},b}},\ldots,e^{\jmath2\pi(Q-1)f_{\text{D},b}}]^T$, and the noise matrix $\mathbf{N}_\text{r}\triangleq[\mathbf{n}_\text{r}[1], \mathbf{n}_\text{r}[2],\ldots,\mathbf{n}_\text{r}[Q]]$. 
Then, the vectorized form $\mathbf{y}_\text{r}\triangleq\text{vec}\{\mathbf{Y}_\text{r}\}$ is given by
\begin{equation}
    \mathbf{y}_\text{r} = \sum_{b=1}^B\alpha_b \mathbf{H}_b\widetilde{\mathbf{S}}\mathbf{w}_b + \sum_{b=1}^B\mathbf{C}_b\widetilde{\mathbf{S}}\mathbf{w}_b + \mathbf{n}_\text{r},
\end{equation}
where for notational simplicity we define
\begin{subequations}
\begin{align}
    \mathbf{H}_b &\triangleq (\mathbf{D}_b\mathbf{J}_{\tau_b}^T) \otimes(\mathbf{a}_\text{r}(\theta_\text{t})\mathbf{a}^T_\text{t}(\theta_{b,\text{t}})),\quad \widetilde{\mathbf{S}} \triangleq \mathbf{S}^T\otimes \mathbf{I}_{N_\text{t}},\\
    \mathbf{C}_b&\triangleq  \mathbf{J}_{\iota_b}^T\otimes\mathbf{G}_b,~~  \mathbf{w}_b = \text{vec}\{\mathbf{W}_b\},~~ \mathbf{n}_\text{r} = \text{vec}\{\mathbf{N}_\text{r}\}.
\end{align}
\end{subequations}
To enhance target detection performance, a space-time receive filter $\mathbf{u} \in\mathbb{C}^{N_\text{r}Q}$ is employed to process $\mathbf{y}_\text{r}$, which yields 
\begin{equation}
   \mathbf{u}^H\mathbf{y}_\text{r} = \mathbf{u}^H\sum_{b=1}^B\alpha_b\mathbf{H}_b\widetilde{\mathbf{S}}\mathbf{w}_b + \mathbf{u}^H\sum_{b=1}^B\mathbf{C}_b\widetilde{\mathbf{S}}\mathbf{w}_b + \mathbf{u}^H\mathbf{n}_\text{r}.
\end{equation}
Thus, the radar output SINR can be calculated as 
\begin{equation}
    \text{SINR}_\text{r} = \frac{\sum_{b=1}^B\sigma^2_b|\mathbf{u}^H\mathbf{H}_b\widetilde{\mathbf{S}}\mathbf{w}_b|^2}{|\mathbf{u}^H\sum_{b=1}^B\mathbf{C}_b\widetilde{\mathbf{S}}\mathbf{w}_b|^2+\sigma^2_\text{r}\mathbf{u}^H\mathbf{u}}.
\end{equation}

In this paper, we propose to jointly optimize the transmit beamforming $\mathbf{W}_b,~\forall b$, and the receive filter $\mathbf{u}$ to maximize the radar output SINR, as well as to satisfy the communication SINR requirements and the power budget. Therefore, the optimization problem is formulated as 
\begin{subequations}\label{eq:original problem}\begin{align}
&\underset{\mathbf{W}_b, \forall b, \mathbf{u}}\max~~\frac{\sum_{b=1}^B\sigma^2_b|\mathbf{u}^H\mathbf{H}_b\widetilde{\mathbf{S}}\mathbf{w}_b|^2}{|\mathbf{u}^H\sum_{b=1}^B\mathbf{C}_b\widetilde{\mathbf{S}}\mathbf{w}_b|^2+\sigma^2_\text{r}\mathbf{u}^H\mathbf{u}}\label{eq:original problem obj}\\
&~~\text{s.t.}~~~\frac{\big|\sum_{b=1}^B\mathbf{h}^T_{b,k}\mathbf{w}_{b,k}\big|^2}{\sum_{j=1,j\neq k}^{K+N_\text{t}}\big|\sum_{b=1}^B\mathbf{h}^T_{b,k}\mathbf{w}_{b,j}\big|^2 + \sigma^2_\text{c}}\geq \Gamma_k,~\forall k, \label{eq:op c1}\\
&\qquad~~~\|\mathbf{W}_b\|_F^2 \leq P_b,~\forall b,\label{eq:op c2}
\end{align}\end{subequations}
where $\Gamma_k$ is the SINR requirement of the $k$-th user, and $P_b$ is the power budget at the $b$-th transmit AP. This non-convex problem is difficult to solve due to the fractional terms and coupled variables. In the next section, we will develop an efficient algorithm to convert problem (\ref{eq:original problem}) into two tractable sub-problems and alternately solve them.

\section{Joint Transmit and Receive Beamforming Design}

In this section, we propose an efficient alternative optimization algorithm for the joint transmit and receive beamforming design problem \eqref{eq:original problem}. The original problem is first decomposed into the receive filter design and transmit beamforming design sub-problems. Then, efficient algorithms are developed to iteratively solve them.

\subsection{Receive Filter Design}

Given fixed transmit beamforming matrix $\mathbf{W}_b,~\forall b$, the receive filter design can be formulated as a generalized Rayleigh quotient, expressed as 
\begin{equation}    \underset{\mathbf{u}}\max~~~\frac{\mathbf{u}^H\sum_{b=1}^B\sigma^2_b\mathbf{H}_b\widetilde{\mathbf{S}}\mathbf{w}_b\mathbf{w}_b^H\widetilde{\mathbf{S}}^H\mathbf{H}^H_b\mathbf{u}}{\mathbf{u}^H\mathbf{R}_\text{i}\mathbf{u}},
\end{equation}
where the covariance of the interference plus noise is defined as 
$\mathbf{R}_\text{i} \triangleq\big(\sum_{b=1}^B\mathbf{C}_b\widetilde{\mathbf{S}}\mathbf{w}_b\big)\big(\sum_{b=1}^B\mathbf{C}_b\widetilde{\mathbf{S}}\mathbf{w}_b\big)^H+\sigma^2_\text{r}\mathbf{I}$. The optimal solution to $\mathbf{u}$ is the eigenvector corresponding to the largest eigenvalue of $\mathbf{T}\triangleq \mathbf{R}_\text{i}^{-1}\sum_{b=1}^B\sigma^2_b\mathbf{H}_b\widetilde{\mathbf{S}}\mathbf{w}_b\mathbf{w}_b^H\widetilde{\mathbf{S}}^H\mathbf{H}^H_b$, which is written as 
\begin{equation}\label{eq:update u}
    \mathbf{u}^\star = \mathbf{v}_\text{max}(\mathbf{T}).
\end{equation}

\subsection{Transmit Beamforming Design}

With the fixed receive filter $\mathbf{u}$, the transmit beamforming design is still a non-convex problem without a closed-form solution due to the complex fractional term in the objective function (\ref{eq:original problem obj}). In order to tackle this issue, we utilize the Dinkelbach's transform and introduce an auxiliary variable $\gamma\in\mathbb{R}$ to convert the fractional term of the objective function (\ref{eq:original problem obj}) into a polynomial expression \cite{Dinkelbach 1967}. Then, the transmit beamforming design problem is reformulated as 
\begin{equation}\label{eq:W transform}\begin{aligned}
&\underset{\mathbf{W}_b, \forall b, \gamma}\max~~\sum_{b=1}^B\sigma^2_b|\mathbf{u}^H\mathbf{H}_b\widetilde{\mathbf{S}}\mathbf{w}_b|^2\\
&\qquad\qquad-\gamma|\mathbf{u}^H\sum\nolimits_{b=1}^B\mathbf{C}_b\widetilde{\mathbf{S}}\mathbf{w}_b|^2-\gamma\sigma^2_\text{r}\mathbf{u}^H\mathbf{u}\\
&\quad~\text{s.t.}\quad~\text{\eqref{eq:op c1}},~\text{\eqref{eq:op c2}}.
\end{aligned}\end{equation}
This bivariate problem can be solved more easily by alternatively updating the auxiliary variable $\gamma$ and the beamforming matrix $\mathbf{W}_b,~\forall b$. 
It is obvious that with fixed $\mathbf{W}_b,~\forall b$, the optimal solution to $\gamma$ is 
\begin{equation}\label{eq:update gamma}
    \gamma^\star = \frac{\sum_{b=1}^B\sigma^2_b|\mathbf{u}^H\mathbf{H}_b\widetilde{\mathbf{S}}\mathbf{w}_b|^2}{|\mathbf{u}^H\sum_{b=1}^B\mathbf{C}_b\widetilde{\mathbf{S}}\mathbf{w}_b|^2+\sigma^2_\text{r}\mathbf{u}^H\mathbf{u}}.
\end{equation}

Given $\gamma$, the objective function is still non-convex due to the first non-convex term in \eqref{eq:W transform}. To obtain a solvable problem for updating $\mathbf{W}_b,~\forall b$, we utilize the idea of majorization-minimization (MM) to construct a convex surrogate function, which approximates the non-convex objective function at the current point and serves as a lower-bound to be maximized in the next iteration \cite{Sun TSP 17}. Specifically, the first-order Taylor expansion is employed to find a convex surrogate function for the term $|\mathbf{u}^H\mathbf{H}_b\widetilde{\mathbf{S}}\mathbf{w}_b|^2$ as  
\begin{equation}\label{eq: surrogate}
|\mathbf{u}^H\mathbf{H}_b\widetilde{\mathbf{S}}\mathbf{w}_b|^2 \geq 2\Re\{\widetilde{\mathbf{w}}_b^H\widetilde{\mathbf{S}}^H\mathbf{H}_b^H\mathbf{uu}^H\mathbf{H}_b\widetilde{\mathbf{S}}\mathbf{w}_b\} - c_b,
\end{equation}
where $\widetilde{\mathbf{w}}_b$ represents the solution to $\mathbf{w}_b$ in the previous iteration and $c_b\triangleq |\mathbf{u}^H\mathbf{H}_b\widetilde{\mathbf{S}}\widetilde{\mathbf{w}}_b|^2$ is a constant term irrelevant to $\mathbf{w}_b$. Based on the result in \eqref{eq: surrogate}, a convex surrogate objective function for \eqref{eq:W transform} is given by 
\begin{equation}
\underset{\mathbf{W}_b,~\forall b}\max~~\sum_{b=1}^B\Re\{\mathbf{f}_b^H\mathbf{w}_b\} - \gamma\Big|\mathbf{u}^H\sum_{b=1}^B\mathbf{C}_b\widetilde{\mathbf{S}}\mathbf{w}_b\Big|^2,
\end{equation}
where we define $\mathbf{f}_b^H\triangleq 2\sigma^2_b\widetilde{\mathbf{w}}_b^H\widetilde{\mathbf{S}}^H\mathbf{H}_b^H\mathbf{uu}^H\mathbf{H}_b\widetilde{\mathbf{S}}$ for simplicity. 
After some matrix transformations, the transmit beamforming design problem can be transformed into a second-order cone programming (SOCP) problem as 
\begin{subequations}\label{eq:update W}\begin{align}
&\underset{\mathbf{W}}\max~~\Re\{\mathbf{f}^H\mathbf{w}\}-|\mathbf{z}^H\mathbf{w}|^2\\
&~~\text{s.t.}~~~(1\!+\! \Gamma_k^{-1})\Re\{\mathbf{g}_k^T\overline{\mathbf{w}}_k\} -\big\|[\mathbf{g}_k^T\mathbf{W},~ \sigma_\text{c}]\big\| \geq 1,~\forall k,\! \\
&\qquad~~~\|\mathbf{T}_b\mathbf{W}\|_F^2 \leq P_b,~\forall b,
\end{align}\end{subequations}
where we define 
\begin{equation}\begin{aligned}
\mathbf{f}^H &\triangleq \sum_{b=1}^B\mathbf{f}_b^H(\mathbf{I}_{N_\text{t}+K}\otimes\mathbf{T}_b),\qquad \mathbf{T}_b \triangleq \mathbf{e}_b^T\otimes \mathbf{I}_{N_\text{t}},\\
\mathbf{W}&\triangleq [\mathbf{W}_1^T,\mathbf{W}_2^T,\ldots,\mathbf{W}_B^T]^T,\qquad \mathbf{w}\triangleq\text{vec}\{\mathbf{W}\},\\
\mathbf{z}^H &\triangleq \sqrt{\gamma}\mathbf{u}^H\sum_{b=1}^B\mathbf{C}_b\widetilde{\mathbf{S}}(\mathbf{I}\otimes\mathbf{T}_b),\\
\mathbf{g}_k^T &\triangleq [\mathbf{h}_{1,k}^T,\ldots,\mathbf{h}_{B,k}^T],~~~ \overline{\mathbf{w}}_k \triangleq [\mathbf{w}_{1,k}^T,\ldots,\mathbf{w}_{B,k}^T]^T,
\end{aligned} \end{equation}
and $\mathbf{e}_b$ is the $b$-th column of the identity matrix $\mathbf{I}_B$. Then, this SOCP problem can be solved using various standard algorithms or tools.

\subsection{Summary}
Based on the above derivations, we summarize the proposed joint transmit beamforming and receive filter design algorithm in Algorithm~1. It is clear that the receive filter $\mathbf{u}$, the auxiliary variable $\gamma$, and the transmit beamforming matrix $\mathbf{W}$ are alternatively updated until convergence. In addition, to obtain a feasible initial solution to $\mathbf{W}$, we maximize the minimum communication SINR under the power budget, 
which is a typical optimization problem that can be converted into an SOCP problem and then easily solved. 

\begin{algorithm}[!t]
\caption{Joint Transmit Beamforming and Receive Filter Design Algorithm.}
\label{alg:1}
    \begin{algorithmic}[1]
    \begin{small}
    \REQUIRE $\mathbf{h}_{b,k}$, $\sigma_\text{c}^2$, $\Gamma_k$, $\sigma_b^2$, $\mathbf{a}_\text{r}(\theta_\text{t})$, $\mathbf{a}_\text{t}(\theta_{b,\text{t}})$, $\tau_b$, $f_{\text{D},b}$, $\mathbf{G}_b$, $\iota_b$, $\sigma^2_\text{r}$, $P_b$.
    \ENSURE $\mathbf{W}_b,~\forall b$, $\mathbf{u}$.
            \STATE {Initialize $\mathbf{W}_b$.} 
            \WHILE {no convergence}
                   \STATE {Update the receive filter $\mathbf{u}$ by \eqref{eq:update u}}
                   \STATE {Update the auxiliary variable $\gamma$ by \eqref{eq:update gamma}.}
                   \STATE {Update the beamforming matrix $\mathbf{W}_b,~\forall b$ by      \eqref{eq:update W}.}
            \ENDWHILE
            \STATE {Return $\mathbf{W}_b,~\forall b$ and $\mathbf{u}$.}
    \end{small}
    \vspace{-0.0 cm}
    \end{algorithmic}
\end{algorithm}

\section{Simulation Results}
\label{Sec:Simulation} 

In this section, we show the simulation results to verify the advantages of the proposed joint space-time adaptive processing and beamforming design for cell-free ISAC systems. We assume that there are $B = 6$ transmit APs and one sensing AP that cooperatively serve $K=15$ users and detect one point-like target with a velocity of 30m/s. Each transmit/receive antenna array is equipped with $N_\text{t}=N_\text{r} = 4$ antennas. The number of symbols transmitted is $L=100$. The carrier frequency is 24GHz, the bandwidth is 10MHz, and the sampling frequency is 20MHz. The channel between the transmitter and the sensing AP/users follows a Rician fading model with a Rician factor of 3dB.
The distance-dependent path-loss exponent for the transmitter-target, target-sensing AP, and transmitter-sensing AP links is 2.2, and for the transmitter-user link is 2.8. The noise power at the receivers is $\sigma^2_\text{r}=\sigma^2_\text{c}=-80$dBm, and the covariance of the target RCS is 1. The communication SINR requirement for each user is the same as $\Gamma_k = \Gamma,~\forall k$. We assume a two-dimensional coordinate system, where the target is located at $(0,0)$, the sensing AP is at $(30, 0)$, the transmit APs are randomly positioned within a ring-shaped area centered on the target with an inner radius of 30m and an outer radius of 60m, and the users are randomly positioned within a circle centered on the target with a radius of 150m. 

\begin{figure}[!t]	\vspace{-0.3 cm} 
\includegraphics[width = \linewidth]{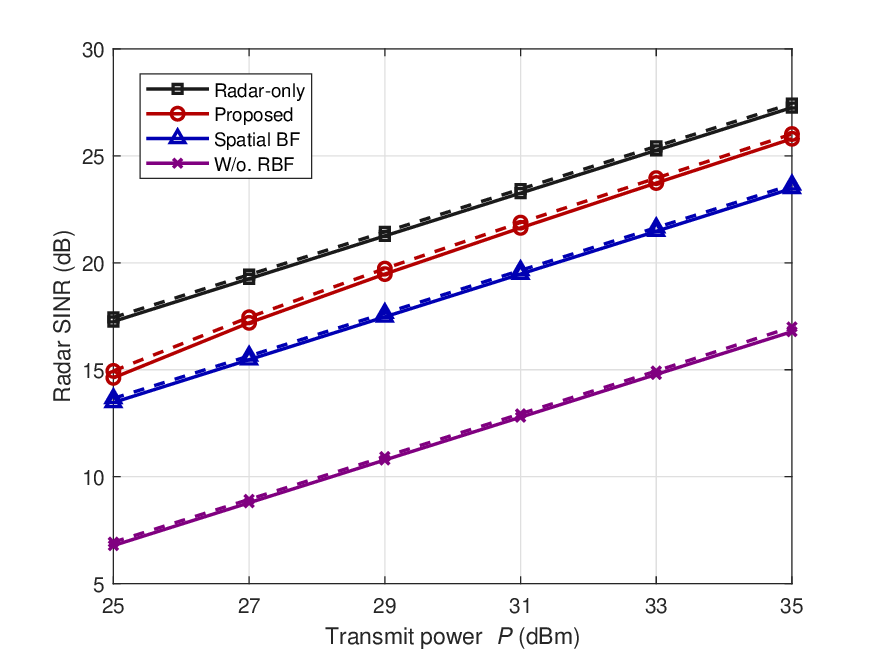}
\vspace{-0.55 cm}
\caption{Radar SINR versus transmit power ($\Gamma=10$dB, dashed lines represent interference links with higher path loss).}\label{fig:power} \vspace{-0.1 cm}
\end{figure} 

We first show the radar SINR versus the transmit power $P = P_b,~\forall b$ in Fig. \ref{fig:power}. The dashed lines represent scenarios where the transmitter-to-sensing AP links experience higher path loss, resulting in reduced clutter interference and improved radar SINR. The proposed joint space-time adaptive processing and beamforming design scheme is denoted as ''\textbf{Proposed}''. For comparison, we include the scheme that designs transmit and receive beamforming in the spatial domain only (''\textbf{Spatial BF}''), the scheme that uses transmit beamforming without receive beamforming (''\textbf{W/o. RBF}''), and the benchmark of the radar-only scenario (''\textbf{Radar-only}''). 
The ``Radar-only'' scheme naturally achieves the highest radar SINR. The proposed scheme incurs a performance loss of nearly 2dB to support a communication SINR of $\Gamma=10$dB for $K=15$ users. 
Compared to ``W/o. RBF'', both the proposed and ``Spatial BF'' schemes achieve significantly higher radar SINR, owing to the benefits of joint transmit and receive beamforming designs. Moreover, compared to ``Spatial BF'', the proposed scheme demonstrates an approximately 2 dB performance improvement, which underscores the advantages of space-time adaptive processing in detecting moving targets, particularly in the presence of strong static clutter.

\begin{figure}[!t]	\vspace{-0.5 cm}
\includegraphics[width = \linewidth]{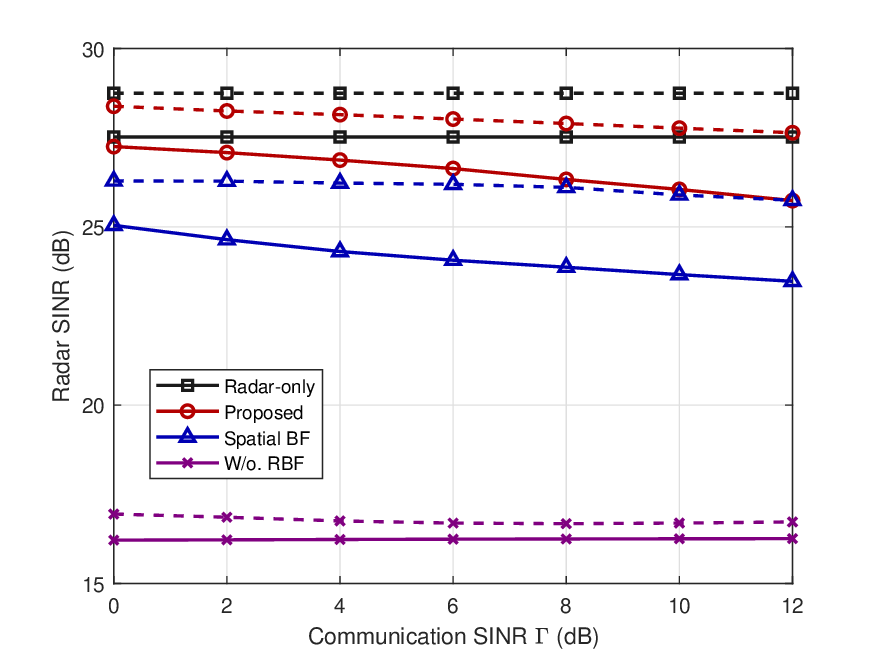}	
\vspace{-0.55 cm}
\caption{Radar SINR versus communication SINR ($P=35$dBm, dashed lines: $N_\text{t}=N_\text{r} = 6$, solid lines: $N_\text{t}=N_\text{r} = 4$).}\label{fig:SINR} \vspace{-0.1 cm}
\end{figure}

Next, radar SINR versus communication SINR requirement $\Gamma = \Gamma_k,~\forall k$ is illustrated in Fig. \ref{fig:SINR}. 
We can clearly observe the performance trade-off between target detection and multiuser communications, as well as the benefits of additional antenna elements in enhancing radar SINR. Moreover, the proposed scheme consistently maintains a higher radar SINR than its counterparts, further confirming the superiority of the proposed joint space-time adaptive processing and beamforming design in cell-free ISAC systems.

\section{Conclusion}\label{sec:conclusion} 
In this paper, we addressed the problem of cooperative target detection and multiuser communications in cell-free ISAC systems. We focused on maximizing the radar SINR for detecting a moving target in the presence of strong clutter, while satisfying the communication SINR constraint and transmit power budget. To address the problem of joint beamforming and receive filter design, we developed an efficient alternative optimization algorithm. Simulation results demonstrated the superiority of the proposed joint space-time adaptive processing and beamforming design in enhancing target detection performance within cell-free ISAC systems.

\end{document}